\documentclass[twoside,fleqn]{article}
\usepackage{espcrc2}
\usepackage{amssymb,amsmath}

\usepackage{graphicx}

\newcommand{\pslash}{\not\!p}
\newcommand{\qslash}{\not\!q}
\newcommand{\Kslash}{\not\!K}
\newcommand{\kslash}{\not\!k}
\newcommand{\half}{\frac{1}{2}}
\newcommand{\quarter}{\frac{1}{4}}

\newcommand{\bra}{\langle}
\newcommand{\ket}{\rangle}
\newcommand{\Tr}{\operatorname{tr}}
\newcommand{\order}{\mathcal{O}}
\newcommand{\MOMB}{$\overline{\text{MOM}}$}
\newcommand{\MOMT}{$\widetilde{\text{MOM}}$}
\newcommand{\MSB}{$\overline{\text{MS}}$}

\newcommand{\gmomt}{g_{\widetilde{\text{MOM}}}}
\newcommand{\amomt}{\alpha_{\widetilde{\text{MOM}}}}

\newcommand{\ams}{\alpha_{\overline{\text{MS}}}}
\newcommand{\Lms}{\Lambda_{\overline{\text{MS}}}}

\newcommand{\Lmomt}{\Lambda_{\widetilde{\text{MOM}}}}
\newcommand{\lmomt}{\lambda_1^{\widetilde{\text{MOM}}}}
\newcommand{\lmomb}{\lambda_1^{\overline{\text{MOM}}}}

\newcommand{\piccie}[2]{\includegraphics[angle=-90,width=#1]{#2}}

\newlength{\colw}
\newcommand{\cquark}{\cite{Skullerud:2000un}}
\newcommand{\cqqg}{\cite{Skullerud:1997wc}}
\newcommand{\cdos}{\cite{Davydychev:2000rt}}
\newcommand{\cgggs}{\cite{Boucaud:1998bq}}

\newcommand{\cLambda}{\cite{Weisz:1995yz,Spitz:1999tu,Capitani:1998mq,Booth:2001qp}}

\title{Quark--gluon vertex in a momentum subtraction
scheme\thanks{Poster presented by J.~Skullerud}}

\author{Jonivar Skullerud\address[DESY]{Theory Division, DESY,
        Notkestra{\ss}e 85, D--22603 Hamburg, Germany},
	Ay{\c s}e K{\i}z{\i}lers\"u\address[CSSM]{Centre for the Subatomic
	Structure of Matter, Adelaide University, Adelaide, SA 5005,
    Australia}, Anthony G.\ Williams\addressmark[CSSM]}

\begin{document}
\makeatletter \@mathmargin = 0pt \makeatother

\begin{abstract}
We compute the quark-gluon vertex in quenched QCD, in the Landau gauge
using an off-shell mean-field $\order(a)$-improved fermion action. The
running coupling is calculated in an `asymmetric' momentum subtraction
scheme (\MOMT).  We obtain a crude estimate for $\Lms=170\pm65$ MeV,
which is considerably lower than other determinations of this
quantity.  However, substantial systematic errors remain.
\end{abstract}

\maketitle

\section{INTRODUCTION}

A nonperturbative study of the quark--gluon vertex is of great
interest for a number of reasons.  Firstly, it allows us to determine
the running coupling $\alpha_s$ from first principles, and also, by
studying the large-momentum behaviour, to determine the scale
parameter $\Lms$.  This approach is complementary to determinations of
$\alpha_s$ from the three-gluon vertex \cgggs, as well as numerous
other methods \cLambda.

Secondly, it may provide input for model studies of hadron structure,
and in particular allow us to assess the reliability of truncation
schemes for Dyson--Schwinger equations.  The infrared behaviour may
also yield information about dynamics of quark confinement
\cite{Alkofer:2000wg}.

Previously \cqqg, the running coupling was studied in an asymmetric
momentum subtraction scheme, and $\order(a)$ errors in the fermion
action were found to be a serious problem.  Here we expand on this
study, using an off-shell $\order(a)$ improved quark propagator to
reduce those errors.

In the continuum, the quark--gluon vertex with gluon momentum $q$ and
quark momenta $p,r=p+q$ can be decomposed as follows: 
\begin{equation} 
\begin{split}
\Lambda_\mu(p,q) & = 
 \sum_{i=1}^{4}\lambda_i(p^2,q^2,r^2)L_{i,\mu}(p,q) \\
 &\phantom{=} + \sum_{i=1}^{8}\tau_i(p^2,q^2,r^2)T_{i,\mu}(p,q)
\end{split}
\end{equation}
\small
The longitudinal components $L_i$ and the transverse
components $T_i$ are given by \cdos
\begin{alignat}{2}
L_{1,\mu} & = \gamma_\mu &\qquad
L_{2,\mu} & = \kslash k_\mu \\
L_{3,\mu} & = k_\mu &\qquad 
L_{4,\mu} & = \sigma_{\mu\nu}k_\nu \notag \\
T_{1,\mu} & = \ell_\mu &\qquad
T_{2,\mu} & = \kslash\ell_\mu \notag\\
T_{3,\mu} & = q^2\gamma_\mu - \qslash q_\mu &\qquad
T_{4,\mu} & = \ell_\mu \sigma_{\nu\lambda}p_\nu q_\lambda \notag\\
T_{5,\mu} & = \sigma_{\mu\nu}q_\nu &\qquad
T_{6,\mu} & = -(qk)\gamma_\mu + \qslash k_\mu \\
T_{7,\mu} & = -\half (qk)\bigl[\kslash&\gamma_\mu - k_\mu\bigr] 
 + &k_\mu\sigma_{\nu\lambda}p_\nu q_\lambda \notag\\
T_{8,\mu} & = -\gamma_\mu\sigma_{\nu\lambda}p_\nu& q_\lambda
 - \pslash q_\mu +& \qslash p_\mu \notag
\end{alignat}
where $k_\mu\equiv(2p+q)_\mu$, $\ell_\mu\equiv (pq)q_\mu-q^2 p_\mu$.  
We are particularly interested in
$\lambda_1$, since this form factor is related to the running
coupling.  In the kinematical limit $q=0$, which we will be
concentrating on here, all the transverse form factors $\tau_i$, as
well as $\lambda_4$, are zero.  We will also be studying the
`symmetric' momentum configuration where $q=-2p$.  In this case, all
the form factors are zero apart from $\lambda_1, \tau_3$ and $\tau_5$.

\section{RENORMALISATION}

We impose `continuum-like' MOM conditions on the quark and gluon
propagators: 
\begin{align}
D^L(qa)|_{q^2=\mu^2} &=
\frac{Z_3(\mu,a)}{\mu^2}  \\
S^L(pq)|_{p^2=\mu^2} &= 
 \left.\frac{Z_2(\mu,a)}{i\Kslash(p)+M(\mu)}\right|_{p^2=\mu^2}
\end{align}
where $K_\mu(p)\equiv\sin p_\mu$.  We then impose momentum subtraction
conditions on $\lambda_1$.  We define the `asymmetric' (\MOMT) scheme by
\begin{equation} 
\lmomt(\mu) \equiv \lambda_1(\mu^2,0,\mu^2)
 = \quarter\Tr\gamma_\nu\Lambda_\nu(p,0)|_{\genfrac{}{}{0pt}{}{p^2=\mu^2}{p_\nu=0}}
\end{equation}
where no sum over the Lorentz index $\nu$ is implied.  It is also
possible to define a `symmetric' (\MOMB) scheme where
$\lmomb(\mu)\equiv\lambda_1(\mu^2,4\mu^2,\mu^2)$; however, as we shall
see it is not possible to implement this scheme in the Landau gauge on
the lattice.

The MOM renormalised coupling is defined by
\begin{equation} 
g_R(\mu) = iZ_2(\mu)Z_3^{1/2}(\mu)\lambda_1(\mu)
\end{equation}


On the lattice, the proper vertex is given by
\begin{multline}
T_{\mu\nu}(q)\Lambda_\nu^a(p,q) = T^aT_{\mu\nu}(q)\Lambda_\mu(p,q) \\ 
 \equiv \bra S(p)\ket^{-1}
\bra S(p)A_\mu^a(q)\ket\bra S(p+q)\ket^{-1}\bra D(q)\ket^{-1}
\label{def:propervtx}
\end{multline}
The tensor $T_{\mu\nu}$ is given by
$D_{\mu\nu}(q)=T_{\mu\nu}(q)D(q)$.  In Landau gauge, for $q\neq0$ this
is simply the transverse projector.  Thus it is not
possible to evaluate the longitudinal components of the quark--gluon
vertex, including $\lambda_1$, for non-zero gluon momentum in Landau
gauge.  This means that our \MOMT\ scheme is the only feasible scheme
in this context.

\section{RESULTS}

We have analysed 495 configurations on a $16^3\times48$ lattice at
$\beta=6.0$, at one quark mass $ma=0.058$, using the
SW action with the mean-field $c_sw=1.479$.  We have used the
`unimproved' quark propagator
\begin{equation}
S_0(x,0) \equiv(M^{-1})_{x0};\quad S_0(p)\equiv \sum_x e^{-ipx}S(x,0)
\end{equation}
and the `improved' propagator \cquark
\begin{equation}
S_I(p) = (1+b_q)S_0(p) + \lambda
\end{equation}
with the mean-field coefficients $\lambda=0.57, b_q=1.14$.  The
configurations have been fixed to the Landau gauge with
$\theta<10^{-12}$.

\begin{figure}
\piccie{\colw}{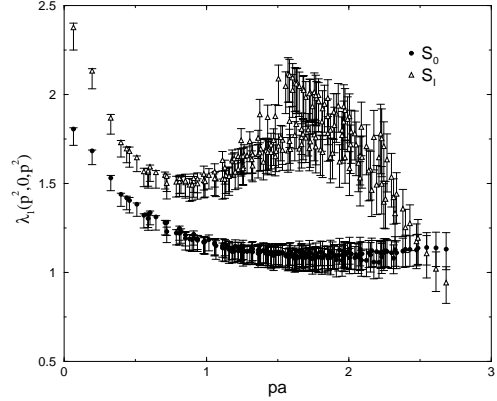}
\vspace{-0.7cm}
\caption{The unrenormalised $\lambda_1(p^2,0,p^2)$ as a function of
$pa$.  The filled circles are data obtained using the `unimproved'
propagator $S_0$, while the open triangles are obtained using the
`improved' propagator $S_I$.}
\label{fig:lambda1}
\end{figure}

In fig.~\ref{fig:lambda1} we show $\lmomt(\mu)$ as a function of
$\mu$, for both $S_0$ and $S_I$.  We see that there is a very big
difference between the unimproved and improved quark propagators.
However, this difference is almost entirely due to the tree-level
behaviour of the improved propagator.  It is possible to implement a
tree-level correction scheme for the vertex similar to the one used
for the quark propagator in \cquark; however, that is not necessary
in this case since the tree-level correction of the vertex is exactly
cancelled by the tree-level correction of $Z_2$ given in \cquark.

\begin{figure}[t]
\piccie{\colw}{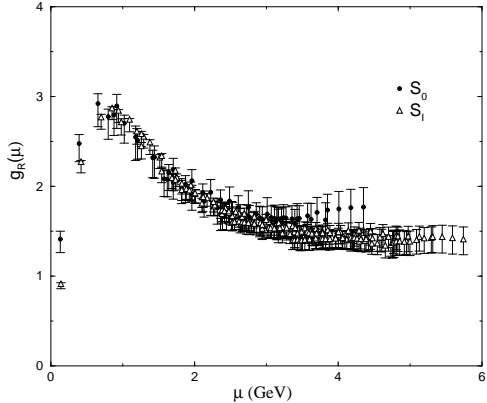}
\vspace{-0.7cm}
\caption{The running coupling $\gmomt(\mu)$ as a function of the
renormalisation scale $\mu$.  The symbols are as in
fig.~\protect\ref{fig:lambda1}.}
\label{fig:gren}
\end{figure}

Fig.~\ref{fig:gren} shows the running coupling $\gmomt(\mu)$ as a
function of $\mu$. We see that the results obtained from $S_0$ and
$S_I$ agree well at intermediate momenta, despite the big difference
in the unrenormalised $\lambda_1$ --- confirming that the dominant
(tree-level) behaviour is cancelled out at the renormalisation stage.
At large momenta, the $S_I$ data are clearly better behaved, whereas
we are not yet in a position to comment on the infrared discrepancy.

We obtain $\Lmomt$ by inverting the two-loop renormalisation
group equation,
\begin{equation}
\Lambda = \mu e^{-\frac{1}{2b_0g^2(\mu)}}
\left(b_0g^2(\mu)\right)^{-\frac{b1}{2b_0^2}}
\label{eq:Lambda}
\end{equation}
\begin{figure}[t]
\piccie{\colw}{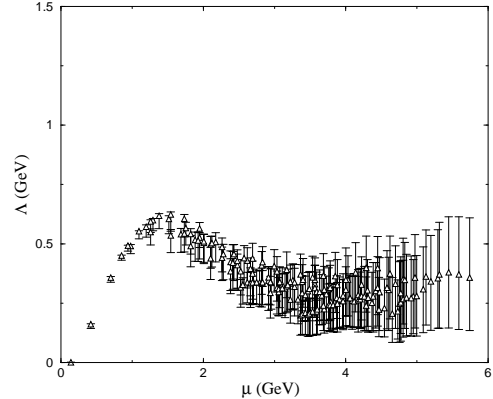}
\vspace{-0.7cm}
\caption{$\Lmomt$ evaluated using (\ref{eq:Lambda}) as function of the
scale $\mu$, using the `improved' propagator $S_I$.}
\label{fig:Lambda}
\end{figure}
The results from $S_I$ are shown in fig.~\ref{fig:Lambda}.  It is not
clear whether there is a perturbative window for this quantity.  We do
not expect two-loop perturbation theory to be valid until $\mu\gg2$
GeV.  It is therefore no surprise that we do not see a plateau in
$\Lambda$ until 3 GeV.  More importantly, as observed in the
three-gluon vertex \cite{Boucaud:2000ey}, $\amomt$ contains power
corrections, which we have not yet taken into account.  We expect the
inclusion of these corrections to significantly change the value of
$\Lmomt$.  On the other hand, lattice artefacts become large for
$\mu>2$ GeV, so it is questionable whether we can trust our data here.

$Z_2, Z_3$ and $\lmomt$ have been computed at one-loop level in the
\MSB\ scheme \cite{Braaten:1981dv,Davydychev:2000rt}.  In Landau
gauge, they are
\begin{align}
Z_3(\mu) &= 1+\frac{\ams(\mu)}{4\pi}\frac{97}{72}C_A \qquad
Z_2(\mu) = 1 \\ 
%
\lmomt(\mu) & =
1+\frac{\ams(\mu)}{4\pi}\frac{C_A}{4}\left[3+\frac{m^2}{\mu^2}\right] 
\label{eq:lambda-pert}
\end{align}
From this we find, for $\mu\gg m$,
\begin{equation}
\amomt(\mu) = \ams(\mu)\left[1\!+\!
\frac{151}{12}\frac{\ams(\mu)}{4\pi}\!+ \order(\alpha^2)\right]
\end{equation}
which gives $\Lmomt/\Lms=\exp(151/232)=1.77$.  From
fig.~\ref{fig:Lambda} we find $\Lmomt=300\pm100$ MeV, which gives an
estimate of $\Lms=170\pm65$ MeV.  As already indicated, however, the
systematic uncertainties connected with the finite lattice spacing
(even using the $S_I$, giving improved ultraviolet behaviour) and the
power corrections to $\alpha_s$ (which come in addition to those
arising in eq.~(\ref{eq:lambda-pert})) are substantial, and may easily
change this estimate by a factor of 2.

\section{OUTLOOK}

We have defined a zero-momentum (\MOMT) subtraction scheme for the
quark--gluon vertex and used it to determine $\alpha_s$ and
$\Lambda_{\text{QCD}}$.  Lattice artefacts still give substantial
uncertainties; it is not clear whether they are under control.  A
further source of systematic error in the determination of
$\Lambda_{\text{QCD}}$ is power corrections to $\alpha_s$.  Work is in
progress to determine these.

In the Landau gauge, longitudinal components of the vertex can only be
studied at zero gluon momentum, so \MOMT\ is the only feasible
renormalisation scheme.  Transverse components, which are all zero at
this point, may be studied in more general kinematics.  We are
currently analysing the two components $\lambda_3(p^2,0,p^2)$ and
$\tau_3(p^2,4p^2,p^2)$. 

In a general covariant gauge, in addition to studying the gauge
dependence of the vertex, it is also possible to define a
symmetric (\MOMB) renormalisation scheme.  This is an interesting
issue for future work.

\section*{Acknowledgments}

This work has been supported by the Australian Research Council and
the TMR-network ``Finite temperature phase transitions in particle
physics'' EU-contract ERBFMRX-CT97-0122.  We thank O.\ P\`ene and R.\
Alkofer for stimulating discussions.


\begin{thebibliography}{10}

\bibitem{Boucaud:1998bq}
P.~Boucaud, {\em et~al.}, 
\newblock JHEP {\bf 10}, 017 (1998); 
%
\newblock hep-ph/0107278.

\bibitem{Weisz:1995yz}
P.~Weisz,
\newblock Nucl. Phys. Proc. Suppl. {\bf 47}, 71 (1996)

\bibitem{Spitz:1999tu}
A.~Spitz {\em et~al.},
\newblock Phys. Rev. {\bf D60}, 074502 (1999)

\bibitem{Capitani:1998mq}
S.~Capitani, M.~Luscher, R.~Sommer, and H.~Wittig,
\newblock Nucl. Phys. {\bf B544}, 669 (1999)

\bibitem{Booth:2001qp}
S.~Booth {\em et~al.},
\newblock hep-lat/0103023.

\bibitem{Alkofer:2000wg}
R.~Alkofer and L.~von Smekal,
\newblock hep-ph/0007355.

\bibitem{Skullerud:1997wc}
J.~I. Skullerud,
\newblock Nucl. Phys. Proc. Suppl. {\bf 63}, 242 (1998)

\bibitem{Davydychev:2000rt}
A.~I. Davydychev, P.~Osland, and L.~Saks,
\newblock Phys. Rev. {\bf D63}, 014022 (2001)

\bibitem{Skullerud:2000un}
%
J.~Skullerud, D.~B. Leinweber, and A.~G. Wil\-liams,
\newblock Phys. Rev. {\bf D64}, 074508 (2001)

\bibitem{Boucaud:2000ey}
P.~Boucaud {\em et~al.},
\newblock JHEP {\bf 04}, 006 (2000)

\bibitem{Braaten:1981dv}
E.~Braaten and J.~P. Leveille,
\newblock Phys. Rev. {\bf D24}, 1369 (1981).

\end{thebibliography}

\end{document}